*Reframing astronomical research through an anticolonial lens — for TMT and beyond*

*Chanda Prescod-Weinstein*, Dept. of Physics & Astronomy and Dept. of Women's & Gender Studies, University of New Hampshire, Durham, NH 03824 (corresponding author: chanda.prescod-weinstein@unh.edu);
*Lucianne M. Walkowicz*, Astronomy Dept., The Adler Planetarium, Chicago IL 60605;
*Sarah Tuttle*, Dept. of Astronomy, University of Washington, Seattle, WA 98103;
*Brian Nord*, KICP, the Dept. of Astronomy & Astrophysics, University of Chicago, Chicago IL 60637;
*Hilding R. Neilson*, Dept. of Astronomy & Astrophysics, University of Toronto, Toronto, ON M5S 3H4

Abstract: This white paper explains that professional astronomy has benefited from settler colonial white supremacist patriarchy. We explicate the impact that this has had on communities which are not the beneficiaries of colonialism and white supremacy. We advocate for astronomers to reject these benefits in the future, and we make proposals regarding the steps involved in rejecting colonialist white supremacy's benefits. We center ten recommendations on the timely issue of what to do about the Thirty Meter Telescope (TMT) on Maunakea in Hawaii. This paper is written in solidarity with and support of efforts by Native Hawaiian scientists (e.g. Kahanamoku et al. 2019).

A central organizing principle of European societies and their settler colonial satellites (e.g. the United States, Canada, and Australia) is that European conceptions of the world and the universe are supreme relative to the way other communities understand it (Whitt 2014; Maile 2015a; Maile 2015b). This phenomenon is widely recognized by humanist and social scientist scholars as settler colonial white supremacist patriarchy (Salazar 2014; Perry 2018). It is often invisible to professional astronomers in so-called western contexts who are typically embedded in this culture from birth (DiAngelo 2018). In this paper, we will refer to this structure using the shorthand of "white supremacy," since the structure relies on both colonialism and patriarchy.

The embedding of white supremacy in European epistemologies manifests in Europe-derived techno-empirical activities which have been professionalized into what we call "professional science" (Whitt 2014). It is therefore the case that the history of science and how it is presently practiced is deeply enmeshed with white supremacist traditions regarding who has been constructed as human, who has been seen as meriting freedom, whose minds are worth cultivating intellectually, and whose relationship(s) to land is upheld and valued (Maile 2015a; Maile 2015b). Importantly, white supremacy specifically benefits people who are welcomed in the tent of whiteness (Prescod-Weinstein 2017), but anyone of any racial and/or ethnic group has the capacity to uphold it as an organizing principle (Kendi 2019).

The impact of white supremacist colonialism on communities that are often referred to as the Global South has been devastating. Walter Rodney's *How Europe Underdeveloped Africa* outlines how Europeans deliberately exploited an entire continent to the point of preventing it from evolving naturally both culturally and environmentally, because it benefited Europe economically to do so (Rodney 1972). European empires, particularly the British, Dutch, French, Belgian, Spanish, and Portuguese, engaged in these kinds of behaviors globally, disrupting indigenous cultures not just in Africa but also across Asia, Polynesia, and throughout North and South America (Loomba 2015).

Science has benefited from these imperial operations in a myriad of ways (see e.g. Prescod-Weinstein 2015 for extensive readings on this topic). One example involves a name quite recognizable to astronomers: Giovanni Domenico Cassini (a.k.a. Jean-Dominique Cassini). In his capacity as director of the Paris Observatory, Cassini sent astronomers to French Guiana to make observations of Mars that, when paired with observations from Paris, allowed them to establish a distance between Earth and Mars (McClellan III 2010). Cassini was also involved in sending astronomers to Saint Domingue (present-day Haiti and the Dominican Republic) to verify that the Earth is not perfectly spherical (McClellan III 2010). NASA has since named a mission after this Cassini (NASA 2019). His son Jacques was involved in a similar expedition involving the Caribbean island of Martinique to look at satellites of Jupiter (McClellan III & Regourd 2000). Importantly, access to all of these places was facilitated by ongoing colonization of the Caribbean and a government interest in better understanding latitudes and longitudes to strengthen the economic benefits of the Atlantic slave trade (Domingues da Silva 2019).

The Caribbean and Latin America are not the only geographies where astronomers intervened in and benefited from colonialism. James Cook was chosen by the British to lead the expedition to observe the Transit of Venus in 1769 as a compromise between the British Navy and the Royal Academy because of his experience observing the 1766 annular eclipse off the coast of Newfoundland (MacLean 1972). As part of the Tahiti expedition, he was given sealed orders to search for Australia, indicating that astronomy and colonization have been entwined in the Pacific since first contact. In the present day, NASA websites continue to celebrate Cook's activities (NASA 2004).

This history is sometimes treated as inevitable and even natural and therefore acceptable. Even if unsavory to thinkers in 2020, it is assumed that people of the time simply couldn't have known any better — but white supremacist patriarchy was never acceptable, and resistance to it existed from its earliest moments. This resistance came in the form of Indigenous self-defense, enslaved people escaping their bondage, and uprisings like the Haitian Revolution (James 1989; Silva 2004; Goodyear-Kaʻōpua, Hussey, & Wright 2014; Dunbar-Ortiz 2015). However, white supremacy has successfully operated as a system that unevenly distributes power: those most likely to object to white supremacist patriarchy have traditionally been locked out of decision-making. It is therefore a problem that the dominant community — people who hold power in this white supremacist context — are still struggling with the question of whether resistance to white supremacy is necessary.

**Resistance to white supremacy is necessary**. It is necessary as a matter of human rights, as a matter of survival in a globally warmed and warming world, and as a matter of making astronomy welcoming to people who have been traditionally excluded from astronomy. This necessity is not a matter of theory but rather one of great practical urgency, particularly in Hawaii, where a long-running struggle over the use of Maunakea has intensified over the last five years (Kahanamoku 2019). Because of the urgency regarding Maunakea, we center recommendations on Maunakea in particular, and we urge readers to join us in recognizing that they are likely applicable and/or adaptable to relationships with other communities as well.

1. Maunakea is part of sovereign Native Hawaiian land, and right of consent to projects like the Thirty Meter Telescope (TMT) belongs to Native Hawaiians. Astronomers have been benefiting from US colonialism in Hawaii for half a century, and Native Hawaiian anti-colonial organizing is now strong enough to take a notable and disruptive stand against these activities on unceded sovereign Native Hawaiian lands (Silva 2004; McGregor & MacKenzie 2014; Salazar 2014; Maile 2015a; Maile 2015b). **We recommend that any astronomers seeking to do work in Hawaii, whether observations or telescope-building, should learn the history and culture of the place and these struggles from the perspective of contemporary traditional cultural knowledge-holders**. **AAS can contribute to this by hosting observatory-funded workshops at their conferences that teach history and culture.**
2. Understand that Native Hawaiians are not a monolith and there is no federal recognition that provides a simplifying organizing principle for "who to listen to." **We advocate acknowledging the particular traditional importance of recognized cultural knowledge-holders. Astronomers must be prepared to meet the challenge of engaging with cultures where authority and decision-making are understood differently from western settler colonial notions of governance.**
3. Because Native Hawaiians (and other indigenous/minoritized communities) are not monolithic, there may be conflict over whether to allow astronomers to proceed with a project or not. Scientists have a duty to evaluate the claims of the parties telling them it's okay and to assess whether those parties fully represent the interests of those who might be most significantly impacted, especially in cases where proceeding might inflict harms that are not easily undone. **Astronomers must recognize past attempts to weaponize differences of opinion in the Indigenous community and commit to not repeating this destructive behavior to achieve a certain outcome. When "contested land" is central to the execution of a project and there is a conflict within an indigenous community about it, astronomers should assess whether moving forward will** **enhance conflict and damage relations between members of indigenous communities. If moving forward will be damaging to such a community, astronomers should pause efforts until consensus can be reached within the community. In light of this, the TMT should halt efforts to build on Maunakea.**

4. Related, communities that are currently operating under colonialism or are in recovery from colonialism are healing from the harms caused by white supremacist colonialism. **We recommend that astronomers consider what is globally healing for the communities rooted in the land, from the perspective of the knowledge-keepers of that community, and they should make an effort to act in ways that are complementary to this healing.**

5. A negotiation is an attempt to reach a mutually acceptable agreement. A good faith negotiation between equals recognizes that one party does not have the power to coerce the other into an agreement. **We recommend that astronomers engage in dialogue and negotiations in good faith, understanding that a deal may not be reachable, with a mandate to respect a "no deal" outcome. Astronomers must accept as part of this process that "consent" has different meanings in different communities, and that these differences are salient to the process (UN Declaration on the Rights of Indigenous People 2007; Neilson et al. 2019; Neilson & Lawler 2019).**

6. Conflict between professionalized astronomers and indigenous communities can and does negatively impact people in astronomy who are rooted in other indigenous communities, including people Indigenous to North America and Black people who have been cut off from their Indigenous African heritages by the slave trade.[1] **We recommend that astronomers be mindful of how their practices globally impact people who have been traditionally marginalized in the astronomy community.**

7. It is tempting to use state authorities to enforce legal rights. Astronomers have rights under settler colonial laws that were designed to override and even deny indigenous rights (Salazar 2014; Maile 2015a; Maile 2015b). "What is legal" and "what is right" do not have a one-to-one mapping in a colonial context. Native Hawaiians, like other indigenous communities, also face disproportionate criminalization under the American criminal justice system. **We recommend astronomers reject the use of state power to get what they want, including any use of force against Native Hawaiians involving police, sheriffs, or the National Guard. Astronomers should also refuse to be party to arrests and prosecutions of Native Hawaiian protectors who violate colonial laws through nonviolent resistance. "Nonviolent" should be broadly interpreted.**

8. Discourse around the TMT and Native Hawaiian resistance to its building has often involved condescending and appropriative language choices by astronomers (Kruesi

---

[1] These groups are not mutually exclusive as there are people who are both Black and enrolled in a Federally recognized tribe or who are otherwise claimed by an Indigenous nation that originates in territories currently governed/occupied by the United States and Canada.

2015). **We recommend that astronomers self-educate about the importance of anti-racism, unlearning anti-indigeneity, and disrupting white fragility (using e.g. DiAngelo 2018; Kendi 2019; Prescod-Weinstein 2015). Further, we recommend that astronomers take a zero-tolerance approach to racist language and cultural appropriation.**

9. Processes like the Decadal Survey on Astronomy and Astrophysics (Astro2020, National Academies 2019) are powerful practices that will determine our professional community's agenda for a decade and beyond. There is now a significant historical record showing that astronomers capitalize on colonial enterprises by building telescopes on lands that have been violently stolen from indigenous peoples. **We recommend that the panel on the State of the Profession give Native Hawaiian Cultural Knowledge-Holders a seat at the table. We recommend that the panel consult with University of Hawaii Dean of Hawaiʻinuiākea School of Hawaiian Knowledge Jonathan Kay Kamakawiwoʻole Osorio about how to implement this recommendation.**

10. Overall, this moment demands a systemic overhaul of the piecemeal manner in which the astronomy community engages with accessing resources, including sites with "good seeing" for astronomical observations. **We recommend initiating the co-generation, with all communities impacted, of ethical guidelines that will underpin astronomical practice ("Kūlana Noiʻi" 2018). Regarding the ethics of site use, the World Archaeological Congress's Code of Ethics (2019) provides a useful example.**

These recommendations might seem overwhelming in the magnitude of the change that they require. They are written in a time when a growing number of students in the astronomical sciences are demanding this kind of accountability and want to be part of a professional community that they can be proud of; adoption of these recommendations is particularly important in creating a practice of astronomy that includes these students, many of whom come from communities that have been traditionally marginalized in astronomy, and who are especially harmed by white supremacy. Data shows that Indigenous people are especially underrepresented in astronomy, and a refusal by astronomers to actively oppose white supremacy can damage their sense of belonging in the community (Neilson et al. 2019).

Helpfully, the Astro2020 is a process by which the community plans for and seeds the foundation of a future we want. By expanding the process to include a consideration of the state of the profession, the community is also laying the foundation for the kind of scientific culture we intend to create. Developing a postcolonial — and even anticolonial — approach to astronomy research is an essential aspect of this process. Adopting this long-overdue approach will change astronomy. "Purely scientific" goals can be more readily achieved when we ignore human rights considerations. Yet, science is a human pursuit, so human rights should be foundational to it.

**Bibliography**

DiAngelo, Robin. *White Fragility: Why It's So Hard for White People to Talk About Racism*. 2018. Boston: Beacon Press.

Domingues da Silva, Daniel B. 2017. *The Atlantic Slave Trade from West Central Africa, 1780-1867*. Cambridge: Cambridge University Press.

Dunbar-Ortiz, Roxanne. 2015. *An Indigenous Peoples' History of the United States*. Boston: Beacon Press.

Goodyear-Kaopua, Noelani, Ikaika Hussey, and Erin Kahunawaika'ala Wright, eds. 2014. *A Nation Rising: Hawaiian Movements for Life, Land, and Sovereignty*. Durham, NC: Duke University Press.

James, C.L.R. 1989. *The Black Jacobins*. Vintage Books ed. New York: Vintage Books.

Kahanamoku, Sara, Hilding Neilson and Rosie 'Anolani Alegado. Forthcoming. "An Indigenous perspective on astronomy on Mauna a Wākea in the time of TMT."

Kendi, Ibram X. 2019. *How to be Antiracist*. New York: One World.

Kruesi, Liz. 2015. "E-mail triggers row over Hawaii telescope." *Physics World*, May 18. https://physicsworld.com/a/e-mail-triggers-row-over-hawaii-telescope/

"Kūlana Noiʻi." 2018. Report. University of Hawaiʻi Sea Grant, June. http://seagrant.soest.hawaii.edu/wp-content/uploads/2018/06/Kulana-Noii-low-res-web.pdf

Loomba, Ania. 2015. *Colonialism/Postcolonialism*. 3rd ed. New York: Routledge.

MacLean, Alistair. 1972. *Captain Cook*. Glasgow: Collins.

Maile, David. 2015a. "Science, Time, and Mauna a Wākea: The Thirty-Meter Telescope's Capitalist-Colonialist Violence, Part I." *The Red Nation*, May 13. https://therednation.org/2015/05/13/science-time-and-mauna-a-wakea-the-thirty-meter-telescopes-capitalist-colonialist-violence-an-essay-in-two-parts/

Maile, David. 2015b. "Science, Time, and Mauna a Wākea: The Thirty-Meter Telescope's Capitalist-Colonialist Violence, Part II." *The Red Nation*, May 20. https://therednation.org/2015/05/20/science-time-and-mauna-a-wakea-the-thirty-meter-telescopes-capitalist-colonialist-violence/

McClellan, James E., III and François Regourd. 2000. "The Colonial Machine: French Science and Colonization in the Ancien Regime." *Osiris* 15:31-50.

McClellan, James E., III. 2010. *Colonialism and Science: Saint Domingue and the Old Regime*. Chicago: University of Chicago Press.

McGregor, Davianna Pōmaikaʻi and Melody Kapilialoha MacKenzie. "Moʻolelo Ea O Nā Hawaiʻi: History of Native Hawaiian Governance in Hawaiʻi." 2014. Report. Office of Hawaiian Affairs, August 19. https://www.doi.gov/sites/doi.opengov.ibmcloud.com/files/uploads/Mo%CA%BBolelo%20Ea%20O%20N%C4%81%20Hawai%CA%BBi(8-23-15).pdf

NASA. "James Cook and the Transit of Venus." May 28, 2004. https://science.nasa.gov/science-news/science-at-nasa/2004/28may_cook/

NASA. "Cassini," Last modified January 28, 2019. https://www.nasa.gov/mission_pages/cassini/main/index.html

National Academies of Sciences, Engineering, and Medicine. "Decadal Survey on Astronomy and Astrophysics (Astro2020)." Accessed November 15, 2019. https://sites.nationalacademies.org/DEPS/astro2020/

Neilson, Hilding, Laurie Rousseau-Nepton, Samantha Lawler, and Kristine Spekkens. 2019. "Indigenizing the next decade of astronomy in Canada." Long Range Plan 2020 White Paper. Canadian Astronomical Society, October 7. https://arxiv.org/abs/1910.02976

Neilson, Hilding and Samantha Lawler. 2019. "Canadian Astronomy on Maunakea: On Respecting Indigenous Rights." Long Range Plan 2020 White Paper. Canadian Astronomical Society, October 8. https://arxiv.org/pdf/1910.03665.pdf

Perry, Imani. 2018. *Vexy Thing: On Gender and Liberation*. Durham, NC: Duke University Press.
7

Prescod-Weinstein, Chanda. 2015. "Decolonising Science Reading List." *Medium*, April 25. https://medium.com/@chanda/decolonising-science-reading-list-339fb773d51f

Prescod-Weinstein, Chanda. 2017. "Black and Palestinian Lives Matter: Black and Jewish America in the Twenty-First Century." In *On Antisemitism: Solidarity and the Struggle for Justice*, edited by Jewish Voice for Peace. Chicago: Haymarket Books.

Rodney, Walter. 1972. *How Europe Underdeveloped Africa*. London: Bogie-L'Ouverture Publications.

Salazar, Joseph. "Multicultural settler colonialism and indigenous struggle in Hawaiʻi : the politics of astronomy on Mauna a Wākea." PhD diss., University of Hawaiʻi at Mānoa, 2014. https://scholarspace.manoa.hawaii.edu/bitstream/10125/101135/1/Salazar_Joseph_r.pdf

Silva, Noenoe K. 2004. *Aloha Betrayed: Native Hawaiian Resistance to American Colonialism*. Durham, NC: Duke University Press.

*United Nations Declaration on the Rights of Indigenous Peoples*, A/RES/61/295 (13 September 07), available from https://www.un.org/development/desa/indigenouspeoples/wp-content/uploads/sites/19/2018/11/UNDRIP_E_web.pdf.

Whitt, Laurelyn. 2014. *Science, Colonialism, and Indigenous Peoples: The Cultural Politics of Law and Knowledge*. Cambridge: Cambridge University Press